\documentclass[%
reprint,
superscriptaddress,
amsmath,amssymb,
aps,
pre,
floatfix,
]{revtex4-2}

\usepackage{graphicx}% Include figure files
\usepackage{bm}% bold math
\usepackage{amsfonts}
\usepackage{siunitx}
\usepackage{xcolor}
\usepackage{physics} % upright d with \dd
\usepackage[normalem]{ulem}
\usepackage{soul}

\usepackage[colorlinks=true,citecolor=blue,urlcolor=red]{hyperref}
\usepackage{etoolbox}
\usepackage{orcidlink}

\DeclareRobustCommand{\demonmark}{%
\raisebox{-0.25ex}{\includegraphics[height=2ex]{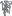}}}

\DeclareRobustCommand{\lyonmark}{%
\raisebox{-0.25ex}{\includegraphics[height=2ex]{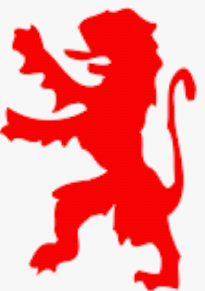}}}

\makeatletter
\def\@fnsymbol#1{%
\ifcase#1\or
\mbox{\demonmark}\or
\mbox{\lyonmark}\or
\ddagger\or
\S\or
\P\or
\|\or
**\or
\dagger\dagger
\else
\@ctrerr
\fi}
\makeatother

\newcommand{\av}[1]{\langle {#1} \rangle}
\newcommand{\kt}{k_B T}
\newcommand{\e}{\mathrm{e}}

\begin{document}
\title{Work as a function of protocol duration for the efficient erasure of an underdamped memory: isothermal to adiabatic transition.}
	
\author{Nicolas Barros\,\orcidlink{0009-0000-1348-7725}}
\affiliation{\href{https://ror.org/02feahw73}{CNRS}, \href{https://ror.org/04zmssz18}{ENS de Lyon}, \href{https://ror.org/00w5ay796}{Laboratoire de Physique}, F-69342 Lyon, France}
\author{Stephen Whitelam\,\orcidlink{0000-0002-0086-6803}}
\email{swhitelam@lbl.gov}
\affiliation{Molecular Foundry, Lawrence Berkeley National Laboratory, 1 Cyclotron Road, Berkeley, CA 94720, USA}
\author{Sergio Ciliberto\,\orcidlink{0000-0002-4366-6094}}
\affiliation{\href{https://ror.org/02feahw73}{CNRS}, \href{https://ror.org/04zmssz18}{ENS de Lyon}, \href{https://ror.org/00w5ay796}{Laboratoire de Physique}, F-69342 Lyon, France}
\author{Ludovic Bellon\,\orcidlink{0000-0002-2499-8106}}
\email{ludovic.bellon@ens-lyon.fr}
\affiliation{\href{https://ror.org/02feahw73}{CNRS}, \href{https://ror.org/04zmssz18}{ENS de Lyon}, \href{https://ror.org/00w5ay796}{Laboratoire de Physique}, F-69342 Lyon, France}
	
\date{\today}

\begin{abstract} 
We use evolutionary reinforcement learning to determine efficient time-dependent erasure protocols for an underdamped cantilever moving in a double-well potential, an experimental realization of a 1-bit memory. We investigate how the mean work $\langle W \rangle $ needed to erase a bit scales as a function of the protocol duration $\tau$. We find two regimes, depending on how $\tau$ compares to the relaxation time of the system $t_{\rm r}$. For $\tau \gg t_{\rm r}$, the quasistatic isothermal regime, we recover Landauer's bound plus an overhead that scales as $1/\tau$, similar to the overdamped case. By contrast, for $\tau<t_{\rm r}$ erasure becomes adiabatic and $\langle W \rangle$ grows more slowly than in the isothermal case. This growth is bounded from below as $1/\tau$, which we derive using a gedanken optimal protocol. Finally, comparison with overdamped erasure shows that learned protocols can outperform protocols that are optimal subject to equilibrium boundary conditions
\end{abstract}
\maketitle
	
\section{Introduction}
Landauer's principle~\cite{landauer1961irreversibility} states that the erasure of a one-bit memory at temperature $T$ requires a minimal work of $k_B T \ln 2$ (where $k_B$ is Boltzmann's constant), reached using a quasi-static process. This prescription stems from about half a century of thought experiments, during which time no experimental tools were available to explore the microscopic systems needed to observe this fundamental limit. However, new experimental techniques, together with the tools offered by stochastic thermodynamics, have in recent decades driven significant progress.

The first experimental verification of Landauer's principle, in 2012, used an overdamped colloidal Brownian particle trapped in a double well potential by optical tweezers~\cite{Berut2012}. In the past decade, other devices such as electrical circuits~\cite{Hong_nano_2016}, feedback traps~\cite{Bech2014}, nanomagnets~\cite{Orlov_2012}, and quantum systems~\cite{QuantumLandauer, SueperconductorLandauer}, have also been used to model memories. Erasure in these devices dissipates energy the order of $k_B T$,  approaching Landauer's bound. In recent years, experiments using optical tweezers in vacuum~\cite{Ciampini-2025} and cantilevers in air~\cite{Dago-2021} have shown that a one-bit memory realized by an {\em underdamped} oscillator presents an interesting option for speed, reliability and energy-efficiency.

All of these experiments have demonstrated that Landauer's bound can be reached with a quasi-static erasure protocol, and that the mean work $\av{W}$ for the 1-bit erasure increases when the protocol duration $\tau$ decreases. The dependence of $\av{W}$ on $\tau$ is not universal: it is determined by the chosen protocol and by the system. However, for overdamped systems described by a first order Langevin equation, the cost of finite-time erasure for overdamped 1-bit memories is well studied. If the end state is forced to be in equilibrium, optimal transport theory proves~\cite{Aurell_2012} that the overhead to Landauer's bound for an optimal erasure protocol of duration $\tau$ scales as
\begin{equation}\label{eq.scaling}
 \langle W\rangle = k_B T \left( \ln 2 + B/\tau \right).
\end{equation}
The constant $B$ is comparable to the free diffusion time over the distance between the 0 and 1 states of the bit, {and can be computed for a given initial double well energy potential~\cite{Aurell_2012,Proesmans-2020,Proesmans-2020-PRE}. A further optimization is possible by allowing an out-of-equilibrium end state, leading to a 4-fold decrease of $B$ for short times $\tau<B$~\cite{Proesmans-2020}.

The $1/\tau$ scaling of Eq.~\ref{eq.scaling}} appears in multiple trade-off relationships governed by out-of-equilibrium  overdamped dynamics~\cite{Klinger2025, Lee2022}. Experimental studies using colloids~\cite{Berut2012, Bech2014, Berut_JSTAT, Proesmans2024, Sagawa2025} and overdamped simulations with a quartic double-well potential~\cite{Eichhorn2023} have verified this result. However, there is no equivalent general proof in the underdamped regime. 
{A bound with a $1/\tau$ scaling can be demonstrated in the case of an equilibrium end state~\cite{Dechant-2019}, but there is no guarantee that this bound can be saturated, and that, in general,} a known applicable optimal protocol to minimize $\av{W}$ exists. For example, recent work in that direction~\cite{sanders2024optimal} does not apply to oscillators with small viscous damping, i.e. with quality factors $Q>1$. Experiments~\cite{Dago-2023-PNAS} have shown that the amount of work for a given erasure time can be reduced by constructing enhanced protocols, following Refs.~\onlinecite{gomez2008optimal, Dellago2010}, in which only specific steps (translations of a single harmonic well) are optimized. These protocols are  not guaranteed to be optimal because only part of the erasure process has been optimized. Thus the question of the dependence on $\tau$ of $\av{W}$ for optimal erasure protocols in underdamped memories is still open. 
 
The purpose of this article is to offer new insight into this problem using machine learning. We encode the erasure protocol as a neural network, and train it using evolutionary methods~\cite{GA, mitchell1998introduction, such2017deep}. This approach, a form of reinforcement learning, is often called neuroevolution. The efficient protocols found by neuroevolution are then implemented in our experimental realization of an underdamped memory~\cite{Dago-2023-PNAS}. The main result of our investigation is that $\av{W}$ for the learned efficient protocols does not follow the simple dependence of Eq.~(\ref{eq.scaling}): this relationship is valid only for $\tau\gg t_{\rm r}$, where $t_{\rm r}$ is the system relaxation time. For fast erasures, i.e. for $\tau\le t_{\rm r}$, Eq.~(\ref{eq.scaling}) does not hold. The second important result is that we propose a gedanken optimal protocol which establishes a bound on the minimum work for fast erasures {in all damping regimes.} We show that the experimental and numerical results of the work are, in our measurement range, always above this bound. {We also directly compare learned protocols in the underdamped and overdamped regimes. We show that, as in the overdamped regime at $\tau<t_{\rm r}$, the work of the learned protocol does not follow the simple scaling of Eq.~\ref{eq.scaling}, and can be smaller than the optimal work when the system ends in equilibrium~\cite{Proesmans-2020-PRE} (a constraint that the learned protocol need not follow).}

We stress that the learned protocols are not guaranteed to be optimal. However, neuroevolution is a robust numerical method that can be applied to all parts of the erasure process, including for rapid driving, and our tests on a range of statistical mechanical systems show that it reliably yields protocols that are optimal or nearly so~\cite{whitelam2023demon, whitelam2025benchmarks,casert2024learning} (see also Refs.~\onlinecite{loos2024universal,szamel2025machine} by other authors). In underdamped systems, for example, optimal protocols can contain features such as delta functions. A neural network cannot represent a delta function exactly, but we find that it can produce regularized approximations that are almost as efficient~\cite{whitelam2025benchmarks,Barros2025,Muratore_2025}.

In the following, we first describe the experimental system used to construct a 1-bit memory. We then explain the neuroevolutionary algorithm used to find efficient erasure protocols, and implement these protocols in laboratory experiments. We compare the work and failure rate of the basic and learned protocols, in simulation and experiment, and show that the mean work exhibits two regimes, determined by the relaxation time of the memory. We then introduce a gedanken optimal protocol that bounds the erasure work, and use it to interpret the fast-erasure data. We end by comparing learned protocols in the underdamped and overdamped regimes.

\begin{figure}[tbp]	
	\centering
	\includegraphics[width=1\columnwidth]{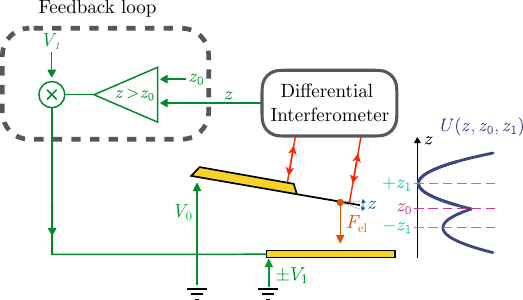}
	\caption{Experimental setup: a conductive cantilever experiences an electrostatic force $F_\mathrm{el}$ due to the voltage difference $V-V_0$ with a facing electrode. The deflection $z$ of the cantilever, measured by interferometry, is compared to a threshold $z_0$ by the feedback loop, setting $V=\pm V_1\ll V_0$. This results in a tunable double-well potential $U(z,z_0,z_1)$~\cite{Dago-2022-JStat,Dago-2024-Chapter}. For a symmetric potential ($z_0 =0$), logical states 0 and 1 are assigned to values $z<0$ and $z>0$, respectively.}
	\label{Fig:set_up} 
\end{figure} 

\section{Experimental setup and basic protocol} 
Fig.~\ref{Fig:set_up} shows our experimental setup. This consists of a micromechanical cantilever that operates, in the absence of external forces, as an underdamped harmonic oscillator characterized by its stiffness $k$, resonance angular frequency $\omega_0 = 2\pi \times (\SI{1090}{Hz})$, and quality factor $Q \approx 7$. The deflection $z$ of the resonator is measured using interferometry~\cite{Paolino-2013}. The oscillator is in thermal equilibrium with the surrounding air at room temperature $T$, and is subject to thermal fluctuations. The variance of $z$ at thermal equilibrium is given by $\sigma^2 = \kt/k \sim \SI{1}{nm^2}$. Subsequently, we express lengths in units of $\sigma$, energies in units of $\kt$, and time in units of the oscillator period $t_0=2\pi/\omega_0$. The relaxation time of this oscillator is $t_{\rm r}= (Q/\pi) t_0 \sim 2 t_0$.
 
Using a fast feedback loop~\cite{Dago-2022-JStat,Dago-2024-Chapter}, we can modulate the electrostatic force acting on the cantilever so that it experiences a virtual bi-quadratic potential energy $U(z,t)$ parameterized by two scalars, $z_0(t)$ and $z_1(t)$: 
\begin{equation} \label{pot}
\begin{split}
 U(z,t) = &\frac{1}{2} \big(z-S[z-z_0(t)]z_1(t)\big)^2\\
 		&+z_0 (t) z_1 (t)\big(S[z-z_0 (t)] +S[z_0 (t)]\big),
\end{split}
\end{equation}
where $S$ is the sign function: $S[z] = - 1$ if $z< 0$, and $S[z] = 1$ otherwise. $U(z,t)$ has in general a double-well form, where $z_0$ tunes the asymmetry and $z_1$ the barrier height. We start and end with a symmetric double well, and associate cantilever positions $z<0$ and $z>0$ with logical states 0 and 1, respectively. Starting in thermal equilibrium, with the cantilever in either well, we can impose a time-dependent protocol $[z_0(t),z_1(t)]$ in order to perform erasure, i.e. to bring the cantilever tip from its starting well to a specified target well in a time $\tau$. In this article, for any given $\tau$, we focus on protocols minimizing the work for a single erasure with a high logical success rate: the final position must correspond to the target well. {We underline that our approach, controlled by only two parameters, is limited to a subset of all possible intermediate energy landscapes. It therefore doesn't allow us to probe fully optimal protocols, that could require more complex non-harmonic potentials~\cite{Sagawa2025}.}

\begin{figure}[tbp]
	\centering
	\includegraphics[width=1\linewidth]{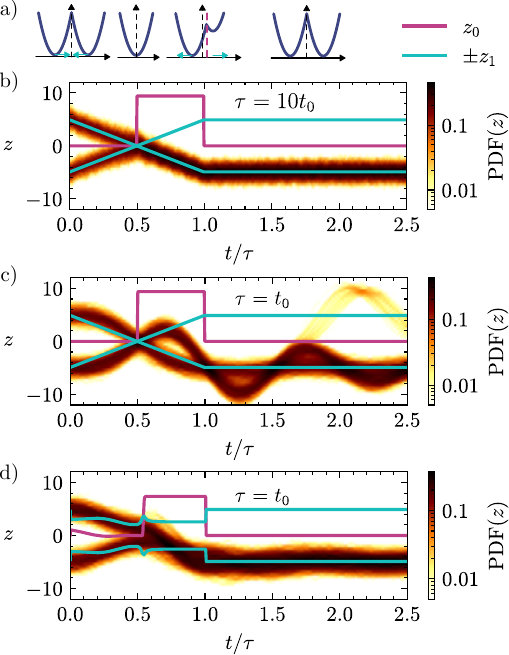}
	\caption{a) Basic erasure protocol for a cantilever evolving in a bi-quadratic double well~\cite{Dago-2021}. The initial symmetric double-well potential, with an energetic barrier of 12.5 $\kt$, encodes a bit of information through the position of the cantilever: 0 if $z<0$, 1 otherwise. In a first stage ($0<t<\tau/2$), the two wells are merged symmetrically, and the information is lost as the cantilever starts jumping between the two states. Then, a translation stage ($\tau/2<t<\tau$) returns the centers of the wells $\pm z_1$ to their original position. This time however, the potential is asymmetric, and a high energetic barrier prevents any transition during the translation. Finally, the initial potential is restored. This linear erasure protocol in a time $t=\tau$ is followed by a resting stage ($t>\tau$). b) For slow erasure ($\tau = 10 t_0$), the basic erasure protocol is efficient as all trajectories end in the target state, here 0 (defined by the bottom half of the plot). c) However, for fast erasure ($\tau = t_0$), some trajectories end up in the wrong state after the erasure operation. d) Conversely, the learned erasure protocol~\cite{Barros2025} achieves a near perfect erasure in the same time. Results presented with the heat maps of panels b)-d) are averages obtained from more than 1000 experimental trajectories corresponding to the superposed protocols.}
	\label{Fig.erasureprotocol} 
\end{figure}

For quasi-static protocols of duration $\tau \gg t_{\rm r}$ (typically $\tau \simeq 100 t_{\rm r}$), we have shown that the Landauer bound can be reached with 1\% uncertainty by following a simple protocol: linear ramps in time from the initial value $Z_1=5$ to $0$ and back for $z_1(t)$, together with large jumps of $z_0(t)$\cite{Dago-2021}. This {\em basic} protocol, pictured in Fig.~\ref{Fig.erasureprotocol}, will serve as a reference throughout our study. Although intuitive and efficient for quasi-static processes, it is neither guaranteed to be optimal nor suitable for all time scales. 

For fast protocols, when $\tau$ is of the order of $t_{\rm r}$, two phenomena negatively affect the operation of underdamped memories. First, residual damping losses are proportional to the velocity, and increase the erasure work. Second, due to its weak coupling to the thermal bath, the system heats up at low dissipation when heat transfer to the thermostat cannot compensate for the work done on the system. The latter energy overhead is described by an extended Landauer bound, where the temperature to consider is a weighted average of $T$ during the transformation~\cite{Dago-2023-PNAS, Dago-2024-APR}. The residual kinetic energy and the heating effect lead to a non-zero {\em failure probability} $P_{\rm f}$, where $P_{\rm f}$ is the fraction of trajectories ending in the wrong well (i.e.~the wrong logical state). The value of $P_{\rm f}$ is measured at a time $t_0$ {\em after} application of the protocol of duration $\tau$ (i.e.~at time $\tau + t_0$, where the quiescent period of duration $t_0$ serves to assess the stability of the logical end state). The question that we analyze in the rest of the article is how the erasure work increases when $\tau$ is reduced, keeping $P_{\rm f}\simeq 0$. As there is no analytical expression for the optimal erasure protocol in an underdamped system, we use neuroevolutionary learning in order to determine protocols that are as efficient as possible. 

\section{Model and learned protocols}
Our model (or ``digital twin'') of the underdamped oscillator consists of the Langevin equation~\cite{Barros2025}
\begin{equation} \label{lang}
 \frac{1}{\omega_0^2}\ddot{z}+ \frac{1}{Q\omega_0}\dot{z} = -U'(z,t)+\sqrt{\frac{2}{Q\omega_0}} \, \xi.
\end{equation}
Here dots and primes denote differentiation with respect to time $t$ and position $z$, respectively, and $\xi$ is a delta correlated Gaussian white noise of unit variance: $\av{\xi(t)}=0$ and $\av{\xi(t)\xi(t')}=\delta(t-t')$, where $\delta$ is the Dirac delta. Introducing the cantilever velocity $v=\dot{z}$ and integrating Eq.~\ref{lang} over the short time interval $\Delta t$ yields the update equations~\cite{Barros2025}
\begin{align}
	z(t+\Delta t)& =z(t)+v(t) \Delta t \label{ex},\\
	v(t+\Delta t)&= \alpha v(t)- \left(1-\alpha\right) Q\omega_0 U'[z(t)] + \omega_0 \sqrt{1-\alpha^2} \xi(t), \nonumber 
\end{align}
where $\alpha \equiv \exp{-\omega_0\Delta t/Q}$. We numerically integrate Eqs.~(\ref{ex}) from time $t=0$ to $t=\tau+t_0$ using a timestep $\Delta t =1.09 \times 10^{-4} t_0$, starting in equilibrium with the potential in its symmetric double-well form.

Reinforcement learning offers a tool for optimization in complex environments, and is well-suited to address physics problems. To produce efficient erasure protocols for a given duration of the protocol $\tau$, we use neuroevolution: we express the protocol $[z_0(t),z_1(t)]$ using a deep neural network, and train it using a genetic algorithm. For efficiency, we choose to apply the learning algorithm to the digital twin of the experiment, Eq.~(\ref{lang}), instead of directly to the experiment. This procedure is detailed in Refs.~\onlinecite{whitelam2023demon,whitelam2025benchmarks}. We instruct the learning algorithm to minimize the order parameter
\begin{equation}\label{eq.singletrain}
 	\phi = P_{\rm f}+ \av{W}/100,
\end{equation} 
where the work $W$ performed by the external electrostatic force is computed for each trajectory in the stochastic thermodynamics framework~\cite{Sekimoto2010},
\begin{equation}
 W=\int_0^\tau \left(\frac{\partial U}{\partial z_0}\dot z_0 + \frac{\partial U}{\partial z_1}\dot z_1\right)dt.
\end{equation}
The mean work $\av{W}$ is evaluated using $10^4$ independent simulated trajectories. The control parameters $z_0(t)$ and $z_1(t)$ are free to take any value during the time $\tau$ of the erasure, and are then held constant (at their default values) for an additional time $t_0$, as in the basic protocol. The value of 100 in the order parameter (\ref{eq.singletrain}) is introduced to make the terms $P_{\rm f}$ and $\av{W}/100$ of similar order of magnitude (and therefore importance) for $P_{\rm f} \lesssim 1\%$. Simulations and experiments are performed by setting the initial positions of the two wells to $\pm Z_1 = \pm 5$, which correspond to an initial energetic barrier $\mathcal{B} = 12.5$.

In our experiments, all protocols are specified as a look-up table of numbers, encoding the values of $z_0(t)$ and $z_1(t)$ for an array of values $t$. This array is computed either analytically, for the basic protocol, or obtained from a neural network trained using neuroevolution. Parameter values are loaded into the digital device controlling the feedback, such that the cantilever experiences a protocol as close as possible to the one numerically defined. The variation of the potential is therefore predetermined, and independent of the cantilever's measured position.

\begin{figure}
	\centering
	\includegraphics[width=1\columnwidth]{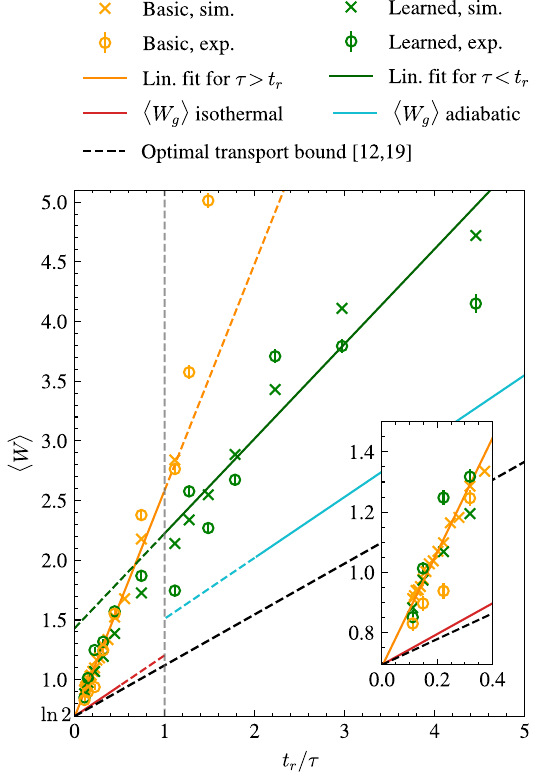}
	\caption{ Underdamped regime - Mean work associated to erasure using the basic and learned
	protocols as a function of $t_{\rm r}/\tau$. We use simulations ($\times$)
	and $\sim 2000$ experimental trajectories ($\circ$, with error bars)
	sampling equally all transitions. We represent a linear fit of the basic
	protocol for long erasures ($\tau>t_{\rm r}$, orange line, fit on simulation data), and of the
	learned protocol for fast erasures ($\tau<t_{\rm r}$, green line, fit on simulation data); each fit
	is extrapolated beyond its range of validity (dashed). The linear fit for $\tau<t_{\rm r}$ presents a large uncertainty on its parameters, due to the large dispersion of the points on the small time range available, and is intended as a guide to the eye. The mean work $\av{W_{\rm g}}$ of the gedanken optimal protocol is shown for the	isothermal (purple) and adiabatic (blue) cases. The black dashed line is the lower bound on the dissipated erasure work set by optimal transport theory~\cite{Dechant-2019, Proesmans-2020-PRE} (see Appendix~\ref{Appendix.Bounds}). The inset provides a zoomed-in view on the long erasures ($\tau \gg t_{\rm r}$).}
	\label{Fig.scaling} 
\end{figure}

\begin{figure}
	\centering
	\includegraphics[width=1\columnwidth]{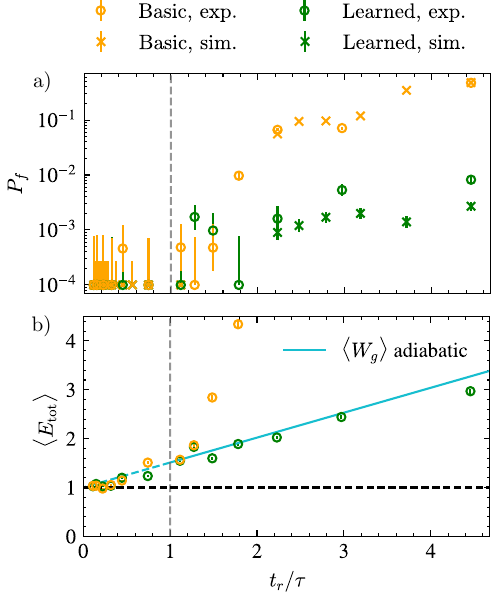}
	\caption{Underdamped regime - a) Failure rate $P_{\rm f}$ as a function of $t_{\rm r}/\tau$ for the single erasure of the learned and the basic protocols. For both protocols $P_{\rm f}<10^{-3}$ for $\tau \gg t_{\rm r}$, but the $P_{\rm f}$ of the basic protocol rapidly increases for shorter $\tau$.
 b) Mean total energy $\langle E_\mathrm{tot} \rangle$ at the end of the erasure protocol ($t=\tau$) as a function of $t_{\rm r}/\tau$. Slower protocols are able to release the accumulated energy as heat during the erasure, and operate isothermally, close to the internal total energy prescribed by the equipartition theorem at equilibrium $\langle E_\mathrm{tot} \rangle= k_B T$. In contrast, fast erasure lead to an increased accumulated total energy. For the basic protocol, the transition between the two regimes appears close to the relaxation time of the system $t_{\rm r}/\tau = 1$. Strikingly, for the learned protocol, $\langle E_\mathrm{tot} \rangle$ coincides with $\langle W_{\rm g} \rangle$, the work performed in the adiabatic gedanken optimal protocol.} 
 	\label{Fig:Ps_v_tau}
	\end{figure}

We have benchmarked the neuroevolutionary learning algorithm across a range of statistical mechanical systems~\cite{whitelam2023demon, whitelam2023train, casert2024learning, whitelam2025benchmarks}. In a recent paper we used the procedure to produce effective and energy-efficient protocols for multiple fast erasures of an underdamped memory~\cite{Barros2025} for a fixed erasure time $\tau = t_0 \sim t_{\rm r}/2$. We found the protocol produced by neuroevolution to be very efficient: it reduces the work of erasure relative to the basic protocol by a factor of 4, and maintains a high success rate, even after multiple successive erasures. In this paper, in order to examine how the erasure work for learned protocols evolves at different erasure times $\tau$, we instruct the learning algorithm to minimize Eq.~\ref{eq.singletrain} for a range $0.5 \ t_0 \leq\tau\leq20 \ t_0$~\footnote{For practical reasons, we limit our exploration range to $\tau \ge t_0/2$ and cannot probe the deep adiabatic regime. Indeed, at shorter time scales, higher-frequency mechanical cantilever modes are excited and can notably affect the fidelity of the feedback loop controlling the potential, causing a divergence from the digital twin. Moreover, the learning algorithm indicates that high-fidelity erasure protocols cannot be found at shorter times {within the limited set of experimentally accessible potentials}. We have therefore restricted the analysis to $t_0/\tau < 2$, which already encompasses erasure protocols considerably faster than the system's relaxation rate.}. We then analyze the performance of both basic and learned protocols numerically and experimentally, using as figures of merit $\av{W}$ in Fig.~\ref{Fig.scaling}, and $P_{\rm f}$ in Fig.~\ref{Fig:Ps_v_tau}.
 
For long erasure times, when $\tau \gg t_{\rm r}$, erasure approaches the quasistatic limit, and basic and learned protocols cost similar amounts of energy (inset of Fig~\ref{Fig.scaling}) and are similarly effective (Fig.~\ref{Fig:Ps_v_tau}). However, for fast erasures ($\tau \lesssim t_{\rm r}$) the learned protocols perform significantly better than the basic protocol. The erasure work (Fig.~\ref{Fig.scaling}) is much smaller for the learned protocol than for the basic, and the failure rate for the learned protocol ranges from $10^{-3}$ to $10^{-2}$ for values of $\tau$ for which the basic protocol fails completely (Fig.~\ref{Fig:Ps_v_tau}). The learned protocols are therefore particularly beneficial for faster erasures ($\tau \leq t_{\rm r}$), where the memory is driven far from equilibrium. The separation between the two regimes occurs when $\tau \sim t_{\rm r}$. 

For long times, where $\tau \gg  t_{\rm r}$, both protocols cost more than the Landauer bound by an amount that grows as $1/\tau$. Fitting the simulations for the basic protocol, we find that in the long-time regime
\begin{equation}
	\langle W \rangle = \ln 2 + B/\tau + C,
    \label{eq:initial_fit}
\end{equation}
with $B = (4.2\, \pm \, 0.1) \, t_0$ and $C = -0.06 \, \pm \, 0.1$. These results are compatible with Eq.~(\ref{eq.scaling}) with $C =0$. 

{Looking at Fig.~\ref{Fig.scaling} we see that  $\av{W}$ for the basic protocol and  $\tau \lesssim t_{\rm r}$ rapidly exceed the initial fit, Eq.~(\ref{eq:initial_fit}).  
Furthermore   for short time  the basic protocol   becomes unreliable,} with $P_{\rm f}$ rapidly exceeding $10^{-1}$ (see Fig.~\ref{Fig:Ps_v_tau}). This behavior coincides with a large increase of the mean total energy measured at the end of the protocol, at $t=\tau$ (Fig.~\ref{Fig:Ps_v_tau}b), which results from the fact that the kinetic energy of the memory increases substantially during the protocol. Because $\tau < t_{\rm r}$, there is no time to dissipate this energy in the heat bath. By contrast, the learned protocol remains very reliable  {for $\tau \lesssim t_{\rm r}$, as $P_{\rm f}$ remains} close to zero. Furthermore, the learned-protocol work (Fig.~\ref{Fig.scaling}) and final energy (Fig.~\ref{Fig:Ps_v_tau}b) grow, as a function of $1/\tau$, much more slowly than the corresponding quantities under the basic protocol. Repeating the numerical study at $Q=14$, we find that, when $Q$ is increased (i.e. when dissipation is reduced), work and energy grow more slowly still (see Appendix~\ref{Appendix.Q}, Fig.~\ref{Fig:compare_Q}).

\section{Gedanken experiment and bound}

Looking at Figs.~\ref{Fig.scaling}, we see that it is difficult to infer the functional form of $\langle W \rangle$ as a function of $\tau$ based only on a fit, because of {the noticeable scatter of the data points and finite time range}. In order to provide a bound on this behavior, we analyze the protocols found by the learning algorithm in terms of our previous experimental and theoretical results.

In Ref.~\onlinecite{Dago-2023-PNAS} we have shown that the work during a translation of a harmonic well can be optimized using the exact results of Refs.~\onlinecite{gomez2008optimal,Dellago2010}. In Appendix~\ref{Appendix.translation}, we show that adding an equilibrium state requirement at the end of the translation on a distance $\Delta Z$ prescribes the optimal work
\begin{equation}
	W^*=\frac{\Delta Z^2}{Q \omega_0 \tau}.
\end{equation}
Furthermore, in underdamped memories, we show that for very fast erasures the transformation switches from isothermal to adiabatic~\cite{Dago-2023-PNAS}. For large quality factors, the heat exchanges with the bath are negligible during the finite-time protocol, and the erasure process boils down to a compression of the volume by a factor of two. This adiabatic compression doubles the system temperature, and requires on average a work of $k_B T$ instead of $k_BT\ln2$, the latter applying only to isothermal protocols.

On the basis of these results let us conceive a gedanken optimal protocol for fast erasures, which fixes a minimum bound for the erasure work. This gedanken reset-to-zero (one) protocol is driven by a demon which reads the initial value of the memory. The demon does nothing if this value is already zero (one), and applies the aforementioned optimal translation if the initial value is one (zero). At time $\tau$, the double well potential is reconstructed at no cost since the system is systematically in the targeted logical state. The minimum mean energy $W_{\rm r}$ needed by the demon for the reading operation is either $1$ (in units of $k_BT$) in adiabatic conditions, or $\ln 2$ in isothermal conditions~\cite{Dago-2023-PNAS} (see Appendix~\ref{Appendix.demon}). Half the time the cantilever starts in the desired final well, in which case $\Delta Z=0$ and $W^*_0=0$. The other half of the time, the cantilever has to be shifted, in which case $\Delta Z=2 Z_1 $ and $W^*_1=4 Z_1^2/(Q\omega_0 \tau)$. Thus the mean total work $\langle W_{\rm g} \rangle$ of the gedanken optimal protocol is
\begin{equation}
	\langle W_{\rm g} \rangle=W_{\rm r}+\frac{W^*_0+W^*_1}{2}=W_{\rm r}+\frac{2Z_1^2}{Q\omega_0\tau}.
	\label{eq:Wg}
\end{equation} 
For large $\tau$, the pertinent limit is isothermal, and Eq.~(\ref{eq:Wg}) takes the form of Eq.~(\ref{eq.scaling}) with a gedanken slope
\begin{equation} \label{EqBg}
B_\mathrm{g}\equiv \frac{2Z_1^2}{Q \omega_0}    
\end{equation}
which evaluates to $B_g\approx1.1t_0$ in our experiment. In the limit of very fast protocols, the $1/\tau$ scaling still holds, except for the intercept at the origin that should reach the adiabatic value of $1$. {$\langle W_{\rm g} \rangle$ is very close to the strict (and possibly unreachable for underdamped systems) lower bound of Ref.~\onlinecite{Dechant-2019} (see Appendix \ref{Appendix.Bounds}). The isothermal gedanken experiment bound is also valid for overdamped systems, and actually closely approaches the optimal protocol bound for an erasure reaching an equilibrium end state (see Appendix \ref{Appendix.Bounds})~\cite{Aurell_2012, Proesmans-2020, Proesmans-2020-PRE}. Out-of-equilibrium end states considered in Ref.~\onlinecite{Proesmans-2020}, that could further decrease the energetic bound, are not considered here: if the initial potential was restored at time $\tau$, they would present a significant failure rate, which we exclude in our approach.}

$W_{\rm g}$ is plotted in Fig.~\ref{Fig.scaling} for $Q=7$, using a transition between $W_{\rm r}=\log 2$ and $W_{\rm r}=1$ at $\tau=t_{\rm r}$. {That bounds are always below the learned protocol, whose results lie close to the adiabatic curve for short $\tau$. The inability for our protocol to reach those bounds is probably rooted in the limited class of potential energy we use in our setup.} Interestingly, as illustrated in Fig.~\ref{Fig:Ps_v_tau}b), $\langle E_\mathrm{tot} \rangle$ coincides with $\langle W_{\rm g} \rangle$ for the adiabatic gedanken optimal protocol.

\section{Comparison of protocols in underdamped and overdamped regimes}

In the previous sections, we have shown that, for underdamped memories, the work of learned erasure protocols does not follow the simple $1/\tau$ law (Eq.~\ref{eq.scaling}) derived for the optimal protocol in overdamped systems. Indeed, we have observed that for $\tau<t_{\rm r}$, the work of the learned protocol grows much more slowly than the $1/\tau$ fit to the data at $\tau>t_{\rm r}$ (Figs.~\ref{Fig.scaling} and \ref{Fig:compare_Q}). We have explained this observation by noting that, for $\tau<t_{\rm r}$, the system enters an adiabatic nonequilibrium regime in which its temperature is higher than that of the bath \cite{Dago-2023-PNAS}. Consequently, the work can be smaller than the optimal work computed under the constraint that the system is in equilibrium at the end of the process (see Ref.~\onlinecite{Proesmans-2020-PRE}).

A natural question is then the following: how does the work of the learned protocols scale in the overdamped regime? In this regime, the optimal work as a function of $\tau$ is known, allowing a direct comparison with the learned protocols. As the focus of this article is on underdamped memories, details of this comparison are given in Appendix~\ref{Appendix.overdamped}; here, we summarize the main results.

Using an overdamped system with the same potential (Eq.~\ref{pot}), we performed three protocols: the basic, learned, and optimal protocols, with the latter computed numerically following Ref.~\onlinecite{Proesmans-2020-PRE} (the latter allows arbitrary variation of the shape of the potential energy, while our learned protocols are constrained to follow the form used in experiments). The results are compared in Fig.~\ref{Fig:overdampedscaling}. We see that, for $\tau<t_{\rm r}$, the work of the learned protocol, which does not impose equilibrium at the end of the process, is smaller than that of the optimal protocol computed with a final-equilibrium constraint. The failure rate of the learned protocol remains very small (Fig.~\ref{Fig:overdampedsuccess}).

We therefore conclude that, as already pointed out in Ref.~\onlinecite{Proesmans-2020-PRE}, in the overdamped regime, protocols that do not impose equilibrium at the end of the process can outperform the optimal protocol under the constraint of equilibrium boundary conditions.

\section{Conclusion}

We used neuroevolutionary learning applied to a simulation model to identify efficient erasure protocols for an underdamped mechanical cantilever. When implemented experimentally, these learned protocols achieve high-fidelity, energy-efficient erasure. Their advantage over hand-designed protocols becomes most apparent for erasure times $\tau$ shorter than the system's relaxation time $t_{\rm r}$. For $\tau > t_{\rm r}$, the mean work follows the quasistatic $1/\tau$ scaling, consistent with overdamped systems. For $\tau < t_{\rm r}$, however, the behavior changes: excess kinetic energy cannot be dissipated into the heat bath within the short erasure time. In this regime, the learned protocols attempt to minimize both the required work and the final stored energy while maintaining near-zero failure rate. The measured and simulated mean work values lie close to the lower bound set by our gedanken optimal protocol, an energetic limit that scales as $1/\tau$ with a coefficient four times smaller than in the slow-erasure regime. This behavior is similar to that observed for overdamped memories, where protocols that do not impose a final equilibrium condition can outperform the optimal protocol computed under this constraint. As noted above, we cannot claim that the learned protocols are optimal, but they provide a clear bound for future studies.

{\em Acknowledgments.} S.C. acknowledges useful discussions with E. Aurell at Nordita. This work has been supported by project ANR-22-CE42-0022. S.W. performed work at the Molecular Foundry, supported by the Office of Science, Office of Basic Energy Sciences, of the U.S. Department of Energy under Contract No. DE-AC02-05CH11231, and partly supported by US DOE Office of Science Scientific User Facilities AI/ML project ``A digital twin for spatiotemporally resolved experiments''. Code for running the simulations can be found in the ``erasure'' folder in the GitHub repository \href{https://github.com/protocol-benchmarks/NESTbench25}{\texttt{protocol-benchmarks/NESTbench25}}. Instructions for training the erasure protocol by neuroevolution can be found in Ref.~\onlinecite{whitelam2025benchmarks}. The data supporting this study will be published in an open repository after acceptance.

\newpage
\appendix

\section*{Appendix}

\section{Comparison of work at different $Q$} \label{Appendix.Q}
\begin{figure}[tbp]
	\centering
	\includegraphics[width=8.5cm]{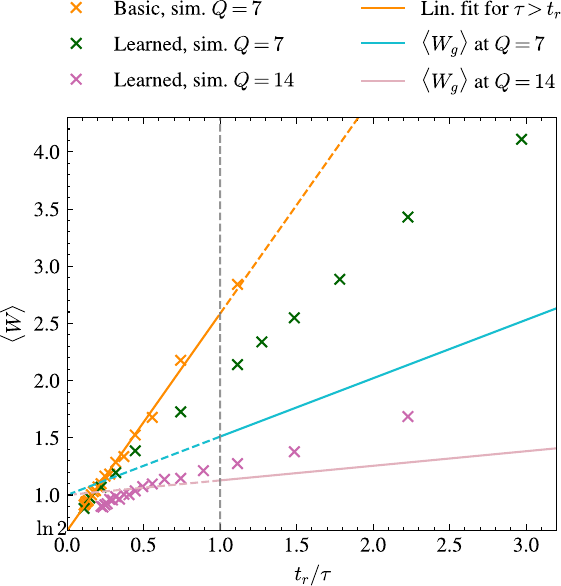}
	\caption{Underdamped regime - Mean work associated to erasure as a function of $t_{\rm r}/\tau$,
	using the basic (orange) and learned (green) protocols computed at $Q=7$ as
	in the main text, and a learned protocol at $Q=14$ (magenta). The mean work
	always lies above the lower bounds predicted by Eq.~\ref{eq:Wg} at $Q=7$
	(light blue line) and $Q=14$ (pink line). The vertical dashed line marks
	$\tau = t_{\rm r}$, common to both quality factors in these rescaled time units.}
	\label{Fig:compare_Q}
\end{figure}

In Fig.~\ref{Fig:compare_Q} we compare the values of $\av{W}$ for the basic protocol at $Q=7$, learned protocols computed at $Q=7$ and $Q=14$, and the corresponding adiabatic bounds for the gedanken protocols $\av{W_{\rm g}}$. We clearly see that these theoretical curves are a bound for the learned protocols. Furthermore we see that when increasing $Q$, the dissipation decreases, {which lowers the bound set by the gedanken protocol. The transition between the long-time $1/\tau$ behavior and the adiabatic one appears at roughly the same rescaled duration, at around $t_{\rm r}/\tau \approx 0.5$ for both quality factors, confirming that the relaxation time $t_{\rm r} = (Q/\pi)\,t_0$ is the relevant time scale separating the two regimes.}
	
\section{Optimal translation protocol with equilibrium end state} \label{Appendix.translation}

Consider the Langevin dynamics of Eq.~\ref{lang} inside a driven harmonic trap $U(z,t)=\frac{1}{2}[z-z_1(t)]^2$:
\begin{equation}
 \frac{1}{\omega_0^2}\ddot{z}+ \frac{1}{Q\omega_0}\dot{z} + z = z_1 + \sqrt{\frac{2}{Q\omega_0}} \, \xi
\end{equation}
This is a linear equation for a simple harmonic oscillator (SHO) with two independent forcing terms: the trap motion $z_1(t)$, and the thermal noise $\xi(t)$. The solution $z=z_D+z_\xi$ is thus the sum of the deterministic motion $z_D$ with no noise and of the thermal noise one in a static harmonic well $z_\xi$, characterized by
\begin{align}
 \frac{1}{\omega_0^2}\ddot{z_D}+ \frac{1}{Q\omega_0}\dot{z_D} + z_D &= z_1(t) \label{eq.z_D}\\
 \av{z_\xi}=0, \av{\dot z_\xi} =0, \av{z_\xi^2} = 1, \av{\dot z_\xi^2} &= \omega_0^2. \label{Eq.z_xi}
\end{align}
The mean heat during a transformation on a time $\tau$ is simply the deterministic one:
\begin{align}
 \av{\mathcal{Q}} & = -\av{\int_0^\tau \frac{\partial U}{\partial z} \dot z dt + \left[\frac{1}{2\omega_0^2}\dot z^2 \right]_0^\tau} \\
 & = \av{\int_0^\tau \left(z_1-z - \frac{1}{\omega_0^2}\ddot z\right) \dot z dt} \\
 & = \int_0^\tau \frac{1}{Q\omega_0}\dot{z_D}^2 dt. \label{Eq.Qintz2}
 \end{align}
Indeed, all terms linear in $z_\xi$ or $\dot z_\xi$ average to zero thanks to Eq.~\ref{Eq.z_xi}, and the quadratic terms correspond to the variation of average total thermal energy $\av{\frac{1}{2}[z_\xi^2+\dot z_\xi^2/\omega_0^2]_0^\tau}=[1]_0^\tau=0$ (alternatively, they can be identified to the heat exchange of a static SHO in equilibrium with a thermostat, which is 0 in average). From Eq.~\ref{Eq.Qintz2}, we deduce the minimum heat lost in the translation
\begin{align}
 \av{\mathcal{Q}} &= \frac{\tau}{Q\omega_0} \frac{1}{\tau}\int_0^\tau \dot{z_D}^2 dt\\
 \av{\mathcal{Q}} &\ge \frac{\tau}{Q\omega_0} \left(\frac{1}{\tau}\int_0^\tau \dot{z_D} dt \right)^2\\
 \av{\mathcal{Q}} &\ge \frac{\Delta Z_D^2}{Q\omega_0\tau}, \label{Eq.Qopt}
 \end{align}
where $\Delta Z_D=z_D(\tau)-z_D(0)$. The bound is saturated when the motion occurs at constant speed $\dot{z_D}=\Delta Z_D/\tau$, which can be recasted into Eq.~\ref{eq.z_D} to compute the necessary driving $z_1(t)$. The jumps in speed [initially from $0$ to $\Delta Z_D/\tau$, finally from $\Delta Z_D/\tau$ to $\dot z_D(\tau^+)]$ are created with Dirac peaks in $z_1$.

In our gedanken experiment, we want not only to minimize the mean work (which is equal to the mean heat for the translation of a harmonic well), but also to secure a low failure rate. We therefore need the cantilever to be far from the barrier between the two wells when restoring the initial potential at time $\tau$, and thus choose to impose $\dot z_D(\tau^+)=0$ and $z_D(\tau)=z_1(\tau)=-z_1(0)$ (the sign depending on the targeted logical state), thus $\Delta Z_D^2 = 4 Z_1^2$. We are then in equilibrium at time $\tau$ in the destination well with a very low probability to cross the barrier (the same as in the static memory), and the optimal translation work is $W^*= 4Z_1^2/(Q\omega_0\tau)$.

Though the preceding bound has been derived in the underdamped case, the result applies to the overdamped realm as well: we just need to remove the acceleration term and kinetic energy ones in all steps of the demonstration (setting $1/\omega_0^2\rightarrow 0$) while rewriting $1/(Q\omega_0)=\gamma/k=t_\mathrm{r}$, with $\gamma$ the damping coefficient and $t_\mathrm{r}$ the relaxation time of the overdamped Langevin equation. The optimal driving $z_1(t)$ does not present Dirac peaks in this overdamped case.

As a final consideration in this appendix, let us consider the case of a non-equilibrium state at time $\tau$. For $t>\tau$, we let the system evolve in a single well centered in $z_1(\tau^+)= - z_1(0^-)$. In the gedanken experiment, we would then delay the restoring of the initial double well potential to later times, after relaxation, as in Ref.\onlinecite{Proesmans-2020}. In such case, the remaining mechanical energy $\frac{1}{2}[(z_D(\tau^+)-z_1(\tau^+))^2+\dot z_D(\tau^+)^2/\omega_0^2]$ of the oscillator is dissipated subsequently, on top of the energy losses during the trap motion expressed by Eq.~\ref{Eq.Qopt}. It leads to the optimal value $\dot z_D(\tau^+)=0$, but it can then be beneficial to aim at $\Delta Z_D<2 Z_1$, the gain during the translation overcoming the potential energy penalty in the final trap. This strategy leads to the optimal translation work in Ref.~\onlinecite{gomez2008optimal}: $\av{W}_\mathrm{non-eq}^*=4Z_1^2/(2+Q\omega_0\tau)$. For example, for infinitely fast translation, it corresponds to directly apply the final potential: the work is finite and corresponds to the increase of potential energy $4Z_1^2/2$, instead of diverging as in Eq.~\ref{Eq.Qopt}. However, the distribution at time $\tau=0^+$ is unchanged with respect to the initial one, so the failure rate would be catastrophic if the final double potential is applied at once.

\section{Comment on the Gedanken optimal protocol} \label{Appendix.demon}
To be precise the demon does not spend energy in the reading: the measurement itself is free. It is instead the subsequent erasure of the demon's memory that costs energy. That is, the demon is carrying out a finite-time operation with the understanding that it, having recorded the initial state of the system, must at some time later erase this information (potentially very slowly).

{\section{Comparison of the Gedanken protocol with bounds from the literature} \label{Appendix.Bounds}

In this appendix, we compare the average work of the gedanken experiment with the bounds of Refs.~\cite{Aurell_2012, Proesmans-2020-PRE, Dechant-2019}. We use here the isothermal $\av{W_\mathrm{g}}$ with $W_\mathrm{r}=\ln 2$, adapted to the comparison with overdamped optimal protocols. In Ref.~\onlinecite{Proesmans-2020-PRE}, we focus only on the protocols ending in an equilibrium state (thus corresponding to the bounds of Refs.~\cite{Aurell_2012, Dechant-2019}), to prevent any failure in the erasure when restoring the initial potential at time $\tau$. These references prescribe that the optimal average work for an overdamped system is $\av{W_\mathrm{opt}}=\ln 2 + B_\mathrm{opt}/\tau$, with
\begin{equation} \label{Bopt}
    B_\mathrm{opt} = \frac{2}{Q\omega_0}\left[\langle z^2\rangle_0
    - \int_0^1 \dd y\, f_0^{-1}(y)\, f_0^{-1}\!\left(\tfrac{y}{2}\right)\right] ,
\end{equation}
where $\av\cdot_0$ is the average evaluated in the initial equilibrium state and $f_0$ is the cumulative function of the initial distribution:
\begin{equation}
    f_0(z) = \frac{1}{\mathcal{Z}_0} \int_{-\infty}^z\dd z'  \e^{-\frac{1}{2}(|z'|-Z_1)^2},
\end{equation}
with $\mathcal{Z}_0$ is the partition function ensuring that $f(+\infty)=1$. These references also compute bounds on $B_\mathrm{opt}$:
\begin{equation} \label{Boptbounds}
    \frac{2}{Q\omega_0}\av{z^2}_0\left(1-\sqrt{1-\frac{\av{|z|}_0^2}{\av{z^2}_0}}\right) \le B_\mathrm{opt} \le  \frac{2}{Q\omega_0}\av{z^2}_0,
\end{equation}
In our bi-quadratic potential centered in $\pm Z_1=5$, we have $\av{|z|}_0=Z_1$ and $\av{z^2}_0=Z_1^2+1$, and using the definition of $B_\mathrm{g}$ in Eq.~\ref{EqBg}, we compute $0.836 \le B_\mathrm{opt}/B_\mathrm{g}\leq 1.04$.  From Eq.~\ref{Bopt}, we can actually compute $B_\mathrm{opt}/B_\mathrm{g}=0.86$: the gedanken performance is close to optimal for the overdamped case. In the underdamped case, no optimal value of the minimum work is known, but the lower bound of Eq.~\ref{Boptbounds} still holds~\cite{Dechant-2019}, illustrating again the pertinence of the gedanken bound $B_\mathrm{g}$, as pictured in Fig.~\ref{Fig.scaling} by the proximity of the purple and dashed black lines.

As a final comment, let us discuss why the optimal protocol can perform better than the gedanken one. To fix ideas, we consider the reset to $0$ erasure, where the final state is a Gaussian distribution of variance $1$ centered in $-Z_1$. The optimal protocol is based on optimal transport, meaning in particular that the half left part of the initial distribution (0 logical state) is transported to the half left part of the final distribution ($z<-Z_1$), while the half right part (1 logical state) is transported to the half right part of the final distribution ($z>-Z_1$). In average, the 0 logical state is translated over a distance $1$, and the 1 logical state over a distance $2Z_1-1$. The optimal translation cost summing those 2 cases would then $\sim[\frac{1}{2}1^2+\frac{1}{2}(2Z_1-1)^2]/(Q\omega_0\tau)=0.82B_\mathrm{g}/\tau$. This order of magnitude demonstrates that a slightly out of equilibrium gedanken protocol could be marginally beneficial, while still ensuring a low failure rate. For the sake of simplicity of the main message of this article, we leave these detailed optimizations aside in the main text.}

{\section{Numerical study of the overdamped erasure} \label{Appendix.overdamped}

\begin{figure}[tbp]
	\centering
	\includegraphics[width=8.5cm]{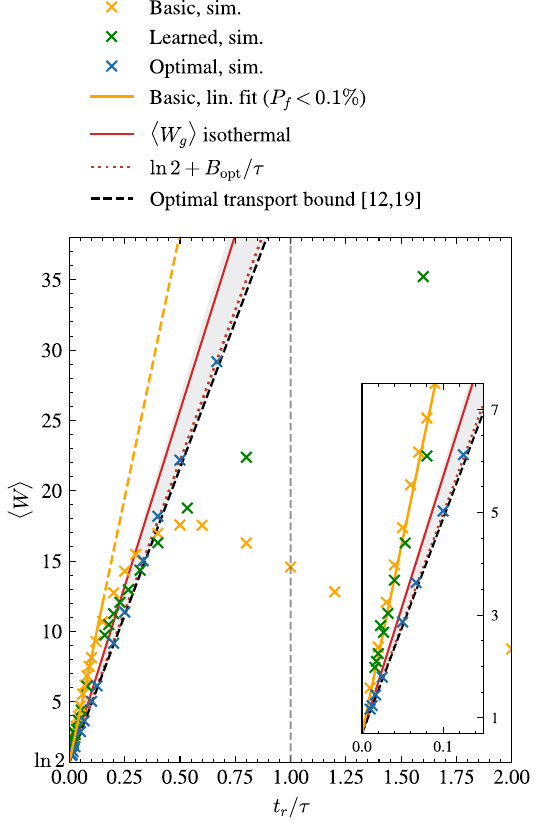}
	\caption{Overdamped regime - Mean work associated to erasure using the optimal, learned and
	basic protocols as a function of $t_{\rm r}/\tau$ in the overdamped regime,
	using $10^4$ simulations. The inset provides a zoomed-in view on the long
	erasures ($\tau \gg t_{\rm r}$), where we show a linear
	fit restricted to highly successful erasures ($P_\mathrm{f} < 0.1\%$) with the intercept fixed at $\ln 2$. The shaded area corresponds to the bounds of
	Eq.~\ref{Boptbounds}, and we provide the analytical curves using
	$B_\mathrm{opt}$ (no fit) and $B_\mathrm{g}$.}
	\label{Fig:overdampedscaling}
\end{figure}

To complement our study, we carry out an analogous analysis in the overdamped regime through numerical simulations, since our experimental setup is restricted to underdamped dynamics. This regime admits an explicit derivation of the optimal erasure protocol for an arbitrary potential, which has been experimentally tested on optical tweezers in Ref.~\onlinecite{Sagawa2025}. Following Ref.~\onlinecite{Proesmans-2020-PRE}, we numerically compute the optimal protocol for an erasure that starts (and ends) at (local) equilibrium in the same biquadratic double well  previously used. The optimal protocol is realized using  a full control of the potential-energy landscape at intermediate times. We compare this optimal protocol with two others, both confined to our experimentally accessible potentials~(\ref{pot}): a protocol learned on an overdamped digital twin, allowing an out-of-equilibrium final state, and the basic linear protocol.

The results of the analysis of the basic, learned and optimal protocols in the overdamped regimes are compared in Figs.~\ref{Fig:overdampedscaling} and~\ref{Fig:overdampedsuccess}. There we plot the mean work $\av{W}$ and the failure rate $P_\mathrm{f}$ as a function of $t_{\rm r}/\tau$ for the three protocols. As already observed when comparing $Q=7$ and $Q=14$ (Fig.~\ref{Fig:compare_Q}) at equivalent protocol time $\tau/t_{\rm r}$, the erasure cost increases as $Q$ decreases; since a larger viscosity corresponds to a smaller $Q$, dissipation is far more important in the overdamped regime, and the underdamped memory studied in Fig.~\ref{Fig.scaling} is more efficient in absolute value for the work.
	
\begin{figure}[b]
	\centering
	\includegraphics[width=8.5cm]{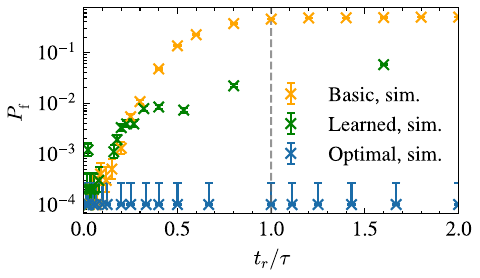}
	\caption{Overdamped regime - Failure rate $P_{\rm f}$ as a function of $t_{\rm r}/\tau$ for the
	single erasure of the basic, learned and optimal protocols in the overdamped
	regime. The optimal protocol, designed analytically to translate the pdf into
	a local-equilibrium state, never fails regardless of $\tau$.}
	\label{Fig:overdampedsuccess}
\end{figure}

For long erasures, in the quasi-static regime shown in the inset, the basic and learned protocols are essentially equivalent, as already observed in the underdamped case. The optimal protocol, benefiting from full control of the potential, can do better: the measured work aligns perfectly with the prediction $\av{W_\mathrm{opt}}=\ln 2 + B_\mathrm{opt}/\tau$ and lies within the optimal-transport bounds of Eq.~\ref{Boptbounds}. Being designed to always attain a local-equilibrium state at $t=\tau$, this protocol is moreover errorless at all $\tau$.

For fast, out-of-equilibrium erasures, however, the learned protocol diverges from the $1/\tau$ scaling at $\tau \simeq 0.5\,t_{\rm r}$ as in the underdamped study. It even becomes more efficient than the optimal protocol under the constraint of equilibrium final condition. As observed in Ref.~\onlinecite{Proesmans-2020-PRE}, relaxing this requirement reduces the erasure cost: while the learned protocol uses only partial control of the potential, it imposes no constraint on the final distribution beyond ending with the correct logical state, which allows it to dissipate less at short times. This crossover occurs around the inflexion at $\tau \simeq 0.5\,t_{\rm r}$, where the learned protocol still presents a high success rate.

Both the underdamped and overdamped regimes thus exhibit the consequences of an
equilibrium / out-of-equilibrium transition near $\tau \simeq t_{\rm r}$. In the
underdamped case, fast erasures leave residual kinetic energy, resulting in an
isothermal-to-adiabatic transition where the final total energy increases along
with the error rate (see Fig.~\ref{Fig:Ps_v_tau}). In the overdamped regime,
where the strong coupling to the thermal bath guarantees isothermal operation,
the possibility of reaching an out-of-equilibrium final distribution lowers the minimal energetic cost.}

\bibliography{UnderdampedErasureWorkvsinvTau}

\end{document}